\documentclass[journal=jpcafh,layout=twocolumn,manuscript=article]{achemso}

\usepackage{graphicx,color}
\pdfoutput=1
\usepackage{amsmath,amsfonts,amssymb}
\usepackage{mathtools}
\usepackage{xspace}
\usepackage{mhchem}
\usepackage{tabularx}
\usepackage{siunitx}
\usepackage{rotating}

\newcolumntype{Y}{>{\centering\arraybackslash}X}

\newcommand{\fig}{Fig.}

\newcommand{\figref}[1]{\fig~\ref{#1}}

\newcommand{\tabref}[1]{table~\ref{#1}}

\renewcommand{\eqref}[1]{Equation~(\ref{#1})}

\newcommand{\ad}[1]{\color{black}{#1~}\color{black}}

\newcommand{\rem}[1]{}%\textcolor{red}{\sout{#1}}}
\newcommand{\stkout}[1]{\ifmmode\text{\sout{\ensuremath{#1}}}\else\sout{#1}\fi}

\newcommand{\chemshell}{ChemShell\xspace}
\newcommand{\dl}{DL-FIND\xspace}

\makeatletter
\def\@dotsep{4.5}
\makeatother
\title{Gaussian process regression for transition state search}
\author{Alexander Denzel}
\author{Johannes K\"{a}stner}
\email{kaestner@theochem.uni-stuttgart.de}
\affiliation{Institute for Theoretical Chemistry, University of Stuttgart, Pfaffenwaldring 55, 70569 Stuttgart,Germany}

\date{\today}
\begin{document}
\begin{abstract}
	We implemented a gradient-based 
	algorithm for transition state search
	which uses Gaussian process regression (GPR).
	Besides a description of the algorithm we provide
	a method to find the starting point for the optimization
	if only the reactant and product minima are known.		
	We perform benchmarks on $27$ test systems
	against the dimer method and partitioned rational function
    optimization (P-RFO) as 
	implemented in the \dl library.
	We found the new optimizer to significantly decrease the number
	of required energy and gradient evaluations.
\end{abstract}
	
\maketitle
	
%%%%%%%%%%%%%%%%%%%%%%%%%%%%%%%%%%%%%%%%%%%%%%%%%%%%%%%%%%%%%%%%%%%%%%%%%%
\section{Introduction\label{sec::Introduction}}
%%%%%%%%%%%%%%%%%%%%%%%%%%%%%%%%%%%%%%%%%%%%%%%%%%%%%%%%%%%%%%%%%%%%%%%%%%
	
\ad{The investigation of reaction mechanisms is a central goal in theoretical
  chemistry. Any reaction can be characterized by its potential energy surface
  (PES) $E(\vec x)$, the energy depending on the nuclear coordinates of all
  atoms. Minima on the PES correspond to reactants and products. The
  minimum-energy path, the path of lowest energy that connects two
  minima, can be seen as an approximation to the mean reaction path. It
  proceeds through a first-order saddle point (SP). Such a SP is the point of
  highest energy along the minimum-energy path. The energy difference between
  a minimum and a SP connected to the minimum is the reaction barrier, which
  can be used in transition state theory to calculate reaction rate
  constants. SPs are typically located by iterative algorithms, with the
  energy and its derivatives calculated by electronic structure methods in
  each iteration. Thus, SP searches are typically the computationally most
  expensive procedures in the theoretical study of reaction mechanisms. Thus,
  efficient black-box algorithms are required to increase the efficiency of
  such simulations. Here, we present such an algorithm based on machine
  learning techniques. 

  A fist-order saddle point is characterized by a vanishing gradient of the
  energy with respect to all coordinates and a single negative eigenvalue of
  the respective Hessian. The eigenmode $\vec{v}_\text{min}$ corresponding to
  that negative eigenvalue is the transition mode, a tangent of the
  minimum-energy path. The SP can be seen as an approximation of the
  transition state. While a general transition state is a surface that
  encapsulates the reactant minimum, the lowest-energy point on such a general
  transition state that minimizes recrossing is in many cases a fist-order
  SP. Thus, a SP is often referred to as transition structure or simply as
  transition state (TS).

  Since the search for transition states is such a central task in
  computational chemistry, many algorithms have been proposed. The most
  general one is probably a full Newton search. This has the disadvantage that
  it converges to any SP, not necessarily first-order ones. Moreover, it
  requires to calculate the Hessian of the potential energy with respect to
  the coordinates at each step, which is computationally rather expensive. An
  algorithm, which also requires Hessian information, but converges specifically
  to first-order SPs is the partitioned rational function
  optimization\cite{Baker_P_RFO} (P-RFO), which is based on the rational
  function approximation to the energy of the RFO method.\cite{RFO} It
  typically shows excellent convergence properties, but its requirement for
  Hessians renders P-RFO impractical in many cases. Algorithms that find
  first-order SPs without Hessian information are the so-called minimum mode
  following methods\cite{MinModeFollowingJonsson1}. They find
  $\vec{v}_\text{min}$ by rotating a dimer on the PES\cite{dimer} or the
  Lanczos method.\cite{Lanczos1950, compareDimerLanczos} By reversing the
  component of the force $\vec{F}$ in the direction of $\vec{v}_\text{min}$
  one can build a new force $\vec{F}^\text{eff} = \vec{F} - 2(\vec{F}\cdot
  \vec{v}_\text{min})\vec{v}_\text{min}$ that takes the algorithms to a
  first-order SP.
%
%Add structure: Intro Methods Results (derzeit Applications) -
%Discussion - Conclusion
%
%Section kriegt eine eigene Zeile, alles darunter ist inline. 
}
	
	\ad{Previous work compared P-RFO and gradient-based 
	minimum mode following methods and found the latter to be advantageous
        in many cases.\cite{compareDimerP_RFO}}	
	Even if they need more steps until convergence, this is compensated by 
	the fact that no Hessians have to be calculated.
	%\ad{It was also shown that it is possible to use GPR to 
	%perform geometry optimization with numerically calculated gradients. \cite{Christiansen}}
	
	The \ad{P-RFO-based
	optimization technique} we present in this paper locates SPs without Hessian information.
	It uses energy and gradient information in 
	the methodology of Gaussian process regression (GPR).\cite{rasmussen2006gaussian}
	\ad{This allows us to use P-RFO on an interpolated PES that is much cheaper 
	to evaluate than the original PES without ever calculating Hessians on the original PES.}
	Kernel-based methods like GPR are increasingly used in theoretical chemistry
	to predict different kinds of chemical properties.
	\cite{Habershon_Tew,Bartok2013,ramakrishnan2017,
		IntramolMultipolKriging,polWaterKriging,
		PredKinEnergyOfCoordsKriging,Ramakrishnan2015,
		hansen2015interaction,PESandVibLevels,
                AtomStrucAmorphousSilicon}
              \ad{Among these are minimization algorithms on the
              PES,\cite{Christiansen,GPRopt} in some cases even without the
              requirement for analytical gradients.\cite{Christiansen}}
	Especially interesting for our case is that GPR was already used to
	search for SPs: 
	the efficiency of the nudged elastic band method (NEB)\cite{MillsNEB_brief, HenkelmanNEB}
	was drastically improved using GPR.\cite{GPR_MEP,GPRNEB_Jonsson}
	In contrast to that approach, we use a surface walking algorithm, focusing on the SP rather than
	optimizing the whole path.
	
	Sometimes it can be difficult to make a good first guess for the
	SP to start the algorithm.
	For the NEB method a procedure was introduced to provide a starting path
	using only geometrical properties of the system
	at two known minima.
	It is called image-dependent pair potential (IDPP).\cite{IDPP}
	If we know the two minima that the wanted TS connects, 
	we will make use of that potential to find an initial guess
	for our TS to start the optimization.

	This paper is organized as follows. 
	We briefly introduce the methodology of GPR in the next section. %\secref{sec::GPR},
	Subsequently we describe in detail how \ad{our optimizations} make use of GPR.
%	In \secref{sec::applications} 
	Then we show some benchmarks of the
	new optimizer and compare it to the well-established
	dimer method and P-RFO in \dl.\cite{dlfind,dimer_dlfind}
	The complete algorithm presented in this paper is
	implemented in \dl and accessible via 
	\chemshell. \cite{SHERWOOD20031,ChemshellReview}
	The code will be made publicly available.
	
	All properties in this paper are 
	expressed in atomic units (Bohr for positions and distances, Hartree for
	energies), unless other units are specified.

%%%%%%%%%%%%%%%%%%%%%%%%%%%%%%%%%%%%%%%%%%%%%%%%%%%%%%%%%%%%%%%%%%%%%%%%%%	
\section{Methods}
\paragraph{Gaussian process regression.\label{sec::GPR}}
%%%%%%%%%%%%%%%%%%%%%%%%%%%%%%%%%%%%%%%%%%%%%%%%%%%%%%%%%%%%%%%%%%%%%%%%%%	
	The idea of GPR is described at length in
	the literature\cite{rasmussen2006gaussian}
	and we will only briefly review the basic idea.
	One can build a surrogate model for the PES
	using $N$ energies $E_1,E_2,...,E_N$ 
	at certain configurations of the molecule 
	$\vec{x}_1,\vec{x}_2,...,\vec{x}_N \in \mathbb{R}^{d}$.
	These configurations are called \emph{training points}.
	In Cartesian coordinates the dimension $d$ of the system is 
	$d=3n$, while $n$ is the number of atoms in the system.
	The key element of the GPR scheme is the covariance
	function $k(\vec{x}_i,\vec{x}_j)$.
	The covariance function describes the covariance
	between two random variables 
	at the points $\vec{x}_i$ and $\vec{x}_j$
	that take on the value of the energies.
	In simplified terms, the covariance function is 
	a similarity measure between these two energies.
	In our case we use a form of the
	Mat\'{e}rn covariance function\cite{matern2013spatial}
	\begin{equation}
	\label{eq::M_Kernel}
	k_{\text{M}}(r)=\left(1+\frac{\sqrt{5}r}{l}+\frac{5r^2}{3l^2}\right)\exp\left[-\frac{\sqrt{5}r}{l}\right]
	\end{equation}
	in which we abbreviated $r=|\vec{x}_i-\vec{x}_j|$.
	The parameter $l$ describes a characteristic length scale 
	of the Gaussian process (GP). It determines how strongly
	the covariance between two random variables (describing energies)
	decreases with distance.
	
	Given a prior estimate $E_{\text{prior}}(\vec{x})$ 
	of the PES, before we have included the training points 
	in the interpolation, one can build the 
	GPR-PES as follows.
	\begin{equation}
		\label{eq::GPprediction}
		E(\vec{x})= \sum_{n=1}^{N}w_n k(\vec{x},\vec{x}_n) + E_{\text{prior}}(\vec{x})
	\end{equation}
	The vector $\vec{w}=(w_1\ w_2\ ...\ w_N)^T$ is the solution of the linear system
	\begin{equation}
		\label{eq::linSystem}
		\sum_{n=1}^{N}K_{mn}w_n=E_m - E_{\text{prior}}(\vec{x}_m)
	\end{equation}
	for all $m=1,2,...,N$.
	The elements
	\begin{equation}
		\label{eq::cov_mat_elements}
		K_{mn}=k(\vec{x}_m,\vec{x}_n)+\sigma_\text{e}^2\delta_{mn}
	\end{equation}
	are the entries of the so called \emph{covariance matrix} $K$ in which
	$\delta_{mn}$ is the Kronecker delta.
	The parameter $\sigma_\text{e}$ describes a noise
	that one assumes on the energy data. 
	If one includes gradients in the scheme, one can introduce an additional
	parameter $\sigma_\text{g}$ for the noise on the gradient data.
	
	Since the kernel function is the only dependency of $x$,
	we can obtain gradients and Hessians of the GPR-PES.
	For the first derivative in dimension $k$, i.e. 
	in the direction of the
	$k$-th unit vector we get
	\begin{equation}
	\frac{dE(\vec{x})}{dx^k}= \sum_{n=1}^{N}w_n \frac{dk(\vec{x},\vec{x}_n)}{dx^k}
	\end{equation}
	and similarly
	\begin{equation}
	\frac{d^2E(\vec{x})}{dx^kdx^l}= \sum_{n=1}^{N}w_n \frac{d^2k(\vec{x},\vec{x}_n)}{dx^kdx^l}
	\end{equation}
	for second derivatives, in dimensions $k$ and $l$.	
	In a previous paper we explicitly describe how one can include
	gradient information into this scheme.\cite{GPRopt}
	In this paper we always use energy and gradient information 
	at the training points.
	In order to build the GPR-PES we then have to solve
	a linear system with a covariance matrix of size
	$N(d+1)$. We solve this linear system via Cholesky decomposition.
	This yields a scaling of $\mathcal{O}\left(N^3d^3\right)$.
	In our case we can decrease the formal scaling to 
	$\mathcal{O}\left(d^3\right)$ 
	with a multi-level approach described below.

%%%%%%%%%%%%%%%%%%%%%%%%%%%%%%%%%%%%%%%%%%%%%%%%%%%%%%%%%%%%%%%%%%%%%%%%%%
\paragraph{The optimization algorithm.\label{sec::optimizer}}
%%%%%%%%%%%%%%%%%%%%%%%%%%%%%%%%%%%%%%%%%%%%%%%%%%%%%%%%%%%%%%%%%%%%%%%%%%	
	
	\ad{In our optimization algorithm} 
	we take a similar approach as the P-RFO optimizer and
	built up a surrogate model for the PES. In contrast to P-RFO we 
	only use energy and gradient information.
	We do not need any second derivatives.
	We use GPR to build the surrogate model by interpolating the energy and 
	gradient information we obtain along the optimization procedure. 
	On the resulting GPR-PES we can cheaply obtain the Hessian
	and therefore, also perform a P-RFO optimization.
	\ad{Our algorithm can be seen as a GPR-based extension
	to P-RFO to dispense the need for Hessian evaluations.}
	The result is used to predict a SP on the real PES.
	We perform all optimizations in Cartesian coordinates.
	
\paragraph{Convergence criteria.}
	We call the vector that points from the last estimate of the TS
	to the current estimate of the TS the \emph{step vector}
	$\vec{s}$, while the gradient at the current estimate of the TS is referred to as
	$\vec{g}$.
	\dl uses a combination of convergence criteria,
	given a single tolerance value, $\delta$.
	The highest absolute value of all entries and the Euclidean norm of 
	both, $\vec{g}$ and $\vec{s}$, have to drop below certain thresholds:
	\begin{alignat}{3}
		\label{eq::convCriteria1}
		\max\limits_i (g_i) &< \delta_{\text{max}(g)} &&\coloneqq \delta \\
		\label{eq::convCriteria2}
		\frac{|\vec{g}|}{d} &< \delta_{|g|} &&\coloneqq \frac{2}{3}&&\delta \\
		\label{eq::convCriteria3}
		\max\limits_i (s_i) &< \delta_{\text{max}(s)} &&\coloneqq 4&&\delta \\
		\label{eq::convCriteria4}
		\frac{|\vec{s}|}{d} &< \delta_{|s|} &&\coloneqq \frac{8}{3}&&\delta
	\end{alignat}	
	\ad{In these equations $d$ stands for the number of dimensions
	in the system.}
	If all these criteria are fulfilled, the TS is considered to be converged.
	These are the same criteria that are used for other TS optimizers
	in \dl.	
	
\paragraph{Parameters.}
	We chose a length scale of $l=20$
	during all optimization runs.
	The noise parameters were chosen as
	$\sigma_\text{e}=\sigma_\text{g}=10^{-7}$.
	\ad{Despite accounting for numerical errors in the electronic
	structure calculations they also function as regularization
	parameters to guarantee convergence of the solution in \eqref{eq::linSystem}.
	We chose the noise parameters as small as possible without compromising
	the stability of the system. We generally found small values for
	the noise parameters to work better for almost all systems.}
	As the prior mean, $E_{\text{prior}}(\vec{x})$,
	of \eqref{eq::GPprediction}, we set the mean value
	of all training points.
	\begin{equation}
	E_{\text{prior}}(\vec{x}) = \frac{1}{N}\sum_{i=1}^{N} E_i
	\end{equation}
%	If we have multiple levels, see our multi-level approach, %\secref{sec::multiLevel},
%	only the level $GP_q$ with the highest index $q$
%	uses this prior. All the others
%	use $GP_{q+1}$ as their prior.
	The parameter $s_{\text{max}}$, the maximum
	step size, must be specified by the user.

\paragraph{Converging the transition mode.}
\label{sec::converge_eigMode}
	We start the optimization at an initial guess $\vec{x}^\text{trans}_0$ 
	for the SP.
	At this point we obtain an approximate Hessian from a GPR-PES constructed from gradient calculations.
	This is done in such a way that we try to converge the eigenvector 
	to the smallest eigenvalue of the Hessian, the transition mode $\vec v$.
	An estimate of that transition mode at the point $\vec{x}^\text{trans}_0$
	is found according to the following procedure, which is the equivalent of dimer rotations in the dimer method.

	\begin{enumerate}
		\item Calculate the energy and the 
		gradient at the point $\vec{x}^\text{trans}_0$
		and also at the point 
		\begin{equation}
			\vec{x}^\text{rot}_1 = \vec{x}^\text{trans}_0+\frac{\Delta}{|\vec{v}_0|}\vec{v}_0
		\end{equation}
		with $\vec{v}_0$ arbitrarily chosen. We generally chose $\vec{v}_0=(1\ 1\ ...\ 1)^T$.
		The results are included as training points in a new GPR-PES.
		We set $\Delta=0.1$ for our optimizer.
		Let $i=1$.
		\item Evaluate the Hessian $H_i(x^\text{trans}_0)$ of the resulting GPR-PES
		at the point $\vec{x}^\text{trans}_0$.
		\item Compute the smallest eigenvalue of $H_i$ and the 
		corresponding eigenvector $\vec{v}_i$.
		\item As soon as 
		\begin{equation}
		\left\lvert\frac{(\vec{v}_i \cdot \vec{v}_{i-1})}{|\vec{v}_i|\cdot |\vec{v}_{i-1}|}\right\rvert>1-\delta_\text{rtol}
		\end{equation}
		we assume that the transition mode is converged, % to the correct eigenvector:
		this procedure is terminated,
		and we move $\vec{x}^\text{trans}_0$ on the PES
		as we describe in the next section.%, see \secref{sec::perform_steps}.
		\item If the transition mode is not converged,
		calculate the energy and gradient at the point 
		\begin{equation}
		\vec{x}^\text{rot}_{i+1} = \vec{x}^\text{trans}_0+\frac{\Delta}{|\vec{v}_i|}\vec{v}_i
		\end{equation}
		and include the results to build a new GPR-PES.
		Increment $i$ by one and go back to step $2$.
	\end{enumerate}
	The procedure to converge the transition mode is not repeated 
	after each movement of $\vec{x}^\text{trans}$ on the PES but only after $50$ such moves.
	In these cases we start at the second 
	step of the described procedure
	with the evaluation of the
	Hessian at the respective point.
	The initial guess for the transition mode after 
	$50$ steps is the vector $\vec{v}_i$ of the last
	optimization of the transition mode.
	We use $\delta_\text{rtol} = 10^{-4}$ 
	(an angle smaller than \ang{0.81}) at the very first point
	$x_0$ and $\delta_\text{rtol} = 10^{-3}$ 
	(an angle smaller than \ang{2.56}) at every
	subsequent point at which we want to converge the transition mode.

\paragraph{Performing steps on the PES.\label{sec::perform_steps}}
	In agreement with the minimum-mode following methods
	we assume that we have enough Hessian-information available
	to move on the PES to the SP
	as soon as the transition mode is found.
	We call the points that are a result from these movements on the PES
	$\vec{x}^\text{trans}_i$, with $i=0, 1, ...$.
	The points $\vec{x}^\text{trans}_i$ correspond to the midpoints in the
	dimer method. 
	We use a user-defined parameter $s_{\text{max}}$ to limit the
	step size along the optimization.
	The step size can never be larger than $s_{\text{max}}$.
	The steps on the PES, starting at the point $\vec{x}^\text{trans}_j$, 
	are performed according to the following procedure.
	\begin{enumerate}
		\item Find the SP on the GPR-PES
		using a P-RFO optimizer.
		This optimization on the GPR-PES is stopped if one of the
		following criteria is fulfilled.
		\begin{itemize}
			\item The step size of this optimization is below $\delta_{\text{max}(s)}/50$.
			\item We found a negative eigenvalue (smaller than $-10^{-10}$) and
			the highest absolute value of all entries of the gradient 
			on the GPR-PES drops below $\delta_{\text{max}(g)}/100$.
			\item The Euclidean distance between the currently estimated TS
			and the starting point of the P-RFO optimization is larger
			than $2s_{\text{max}}$.
		\end{itemize}
		If none of these are fulfilled after $100$ P-RFO steps,
		we use a simple dimer translation. The dimer translation is
		done until one of the criteria above is fulfilled
		or the  Euclidean distance between the currently estimated TS
		and the starting point is larger
		than $s_{\text{max}}$.
		The P-RFO method converged in less than $100$ steps 
		in all of the presented test cases of this paper.
		\item Overshooting the estimated step, according to the
		overshooting procedures described in %\secref{sec::overshooting}
		the next section, resulting in $\vec{x}^\text{trans}_{j+1}$.
        \item Calculate the energy and gradient at 
        $\vec{x}^\text{trans}_{j+1}$ and build a new GPR-PES.
	\end{enumerate}
	The used P-RFO implementation for this procedure
	is the same as used in \dl (with Hessians of the GPR-PES re-calculated in each step).
	P-RFO tries to find the modes corresponding to overall
	translation and rotation of the system and ignores them.
	Numerically, it is not always clear 
	which modes these are.
	Therefore, we found it to be very beneficial to project 
	the translational modes out of the inferred Hessian.
	This yields translational eigenvalues that are numerically zero.
	Otherwise, the TS search via P-RFO tends to translate the
	whole molecule, leading to an unnecessary large step size.

\paragraph{Overshooting.}
	\label{sec::overshooting}
	As we have done in the optimization algorithm for
	minimum search,\cite{GPRopt}
	we try to shift the area in which we predict the
	SP to an interpolation regime
	rather than an extrapolation regime.
	This is because machine learning methods perform poorly
	in extrapolation.
	The overshooting is done, dependent on the
	angle between the vectors along the optimization:
	let $\vec{s}_{N}^{\, \prime}$ be the vector from the
	point $\vec{x}^\text{trans}_{N-1}$ to the next estimate for the TS 
	according to the first step in the procedure described in the previous section. %of \Secref{sec::perform_steps}.
	Let $\vec{s}_{N-1}$ be the vector pointing from 
	$\vec{x}^\text{trans}_{N-2}$ to $\vec{x}^\text{trans}_{N-1}$.
	If $\vec{x}^\text{trans}_{N-1}=\vec{x}_0$, the point $\vec{x}^\text{trans}_{N}$ 
	is simply calculated according to the procedure described in the previous section
%	\Secref{sec::perform_steps} 
	with no overshooting.
	We calculate an angle
\begin{equation}
	\label{eq::angle}
	\alpha_N = \frac{(\vec{s}_{N-1}\cdot \vec{s}^{\, \prime}_N)}{|\vec{s}_{N-1}||\vec{s}^{\, \prime}_N|}
\end{equation}
	and introduce a scaling factor 
\begin{equation}
	\label{eq::overshooting}
	\lambda(\alpha_N) = 1+ ({\lambda}_{\text{max}}-1)\left(\frac{\alpha_N-0.9}{1-0.9}\right)^4
\end{equation}
	that scales
\begin{equation}
	\vec{s}_N={\lambda}(\alpha_N)\vec{s}^{\, \prime}_N
\end{equation}
	as soon as $\alpha_N > 0.9$.
	The scaling limit $\lambda_{\text{max}}$ is chosen to be $5$.
	But it is reduced if the algorithm is close to convergence and
	it is increased if $\alpha_N > 0.9$ for two 
	consecutive steps, i.e. we overshoot more than once
	in a row. We refer to our previous work for a more
	detailed description of the calculation 
	of $\lambda_{\text{max}}$.\cite{GPRopt}
	We also use the \emph{separate dimension overshooting}
	procedure from that paper:
	if the value of a coordinate along the steps
	$\vec{x}^\text{trans}_i$ changed monotonically for the last $20$ steps, 
	a one dimensional
	GP is used to represent the development of this coordinate
	along the optimization. It is optimized, independent
	from the other coordinates.
	To account for coupling between the coordinates in succeeding
	optimization steps this procedure is suspended
	for $20$ steps after it was performed.
	After these two overshooting procedures the step is
	scaled down to $s_{\text{max}}$.
	The chosen overshooting procedures might seem
	too aggressive at first glance. 
	But the accuracy of the GPR-PES is 
	largely improved if we overshoot the real
	TS. This is a crucial difference to 
	conventional optimizers:
	a step that is too large/in the wrong direction
	can still be used to improve the next estimate of the TS.

\paragraph{Multi-level GPR.}
	\label{sec::multiLevel}
	Just like in our previous work\cite{GPRopt}
	we include a multi-level scheme to reduce the computational
	effort of the GPR.
	\ad{A more detailed explanation of the multi-level approach
	can also be found there.
	The most demanding computational step in the GPR scheme 
	is solving \eqref{eq::linSystem}.
	The computational effort to solve this system 
	can be reduced by limiting the size of the covariance matrix. 
	Our multi-level scheme achieves this
	by solving several smaller systems
	instead of one single large system.
	This eliminates} the formal scaling with 
	the number of steps in the optimization history.
	To achieve this, we take the oldest
	$m$ training points in the GP to build a separate
	GP called $GP_1$. This is done as soon as the number of
	training points reaches $N_{\text{max}}$.
	The predicted GPR-PES from $GP_1$ is used as $E_{\text{prior}}(\vec{x})$ in \eqref{eq::GPprediction}
	for the GP \ad{built with} the remaining $N_{\text{max}}-m$ training points
	that we call $GP_0$.
	The new training points are added to $GP_0$.
	When $GP_0$ eventually has more than $N_{\text{max}}$ training points 
	we rename $GP_1$ to $GP_2$ and use the $m$ oldest
	training points in $GP_0$ to build a new $GP_1$.
	We always take $GP_{i+1}$ as the prior to $GP_i$.
	The $GP_q$ with the highest index $q$ just uses
	the mean of all energy values at the contained
	training points as its prior.
	The number of levels increases along the optimization
	but since the number of points in all $GP_i$ is
	kept below a certain constant the formal scaling of
	our GPR scheme is $\mathcal{O}\left(d^3\right)$.
	\ad{Splitting the points which were used to converge the
	smallest eigenvalue of the Hessian (according to
	the procedure for the converging of the transition mode
	described above) leads to a loss in accuracy of the
	second order information of the GPR-PES and 
	consequently, to inaccurate
	predictions of the step direction by the underlying P-RFO.}
%	We found that it is beneficial not to split 
%	the points which were used to converge the
%	smallest eigenvalue of the Hessian according to
%	the procedure for the converging of the transition mode
%	described above.%of \Secref{sec::converge_eigMode}.
	Therefore, we increase $N_\text{max}$ and $m$ by
	one if those points would be split into 
	different levels and try the splitting again
	after including the next training point.
	When we successfully performed the splitting,
	the original values of $N_\text{max}$ and $m$ are
	restored.
	We set the values $N_\text{max}=60$ and
	$m=10$ for all tests performed in this paper.
	We also want to guarantee that we always have
	information about the second derivative 
	in $GP_0$. Therefore, we start
	our Hessian approximation via
	Procedure 1 after $N_\text{max}-m = 50$ steps on the PES.
	As a result, the points that belong to the last
	Hessian approximation are
	always included in $GP_0$.
    The method described so far is referred to as GPRTS in the following.

\paragraph{Finding a starting point via IDPP.\label{sec:: GPRPP}}

	If one intends to find a TS that connects two
	known minima, the algorithm tries to find
	a good guess for the starting point of the TS search
	automatically.
	To that end we optimize a nudged elastic band (NEB)
	\cite{MillsNEB_brief, HenkelmanNEB}
	in the image dependent pair potential (IDPP).\cite{IDPP}
	The construction of this potential does not need
	any electronic structure calculations,
	but is a good first guess for a NEB path 
	based on geometrical considerations.
	Given $M$ images $\vec{x}^\text{IDPP}_i,\ i = 1, ..., M$
	of the optimized NEB in the 
	IDPP, our algorithm calculates
	the energy and gradient of the real PES
	at image $\vec{x}^\text{IDPP}_j$ with $j={M/2}$ for even 
	and $j=(M-1)/2$ for odd $M$.
	\ad{Only from this point on we perform energy evaluations
	of the real PES.}
	Then we calculate the scalar product between the gradient
	$\vec{\gamma}_j$
	at point $\vec{x}^\text{IDPP}_j$ and the vector
	to the next image on the NEB path.
\begin{equation}
	\beta = \vec{\gamma}_j\cdot (\vec{x}^\text{IDPP}_{j+1}-\vec{x}^\text{IDPP}_j)
\end{equation}
	If $\beta>0$, we calculate
	the energy and the gradient at $\vec{x}^\text{IDPP}_{j+1}$, otherwise at $\vec{x}^\text{IDPP}_{j-1}$ in order to aim at a maximum in the energy.
	This is repeated until we change the direction along 
	the NEB and would go back to a point $\vec{x}^\text{IDPP}_{\text{best}}$ at which
	we already know the energy and the gradient.
	The point $\vec{x}^\text{IDPP}_{\text{best}}$ we found in this procedure
	is assumed to be the best guess for the TS on the NEB and is
	the starting point for our optimizer.
	All the energy and gradient information acquired
	in this procedure is used to build the GPR-PES. 
	The optimization technique including the 
	usage of the NEB in the IDPP potential
	will be abbreviated as \emph{GPRPP}.

%%%%%%%%%%%%%%%%%%%%%%%%%%%%%%%%%%%%%%%%%%%%%%%%%%%%%%%%%%%%%%%%%%%%%%%%
\section{Results\label{sec::applications}}
\begin{figure}%[h!]
		\begin{center}
			\includegraphics[width=\linewidth]{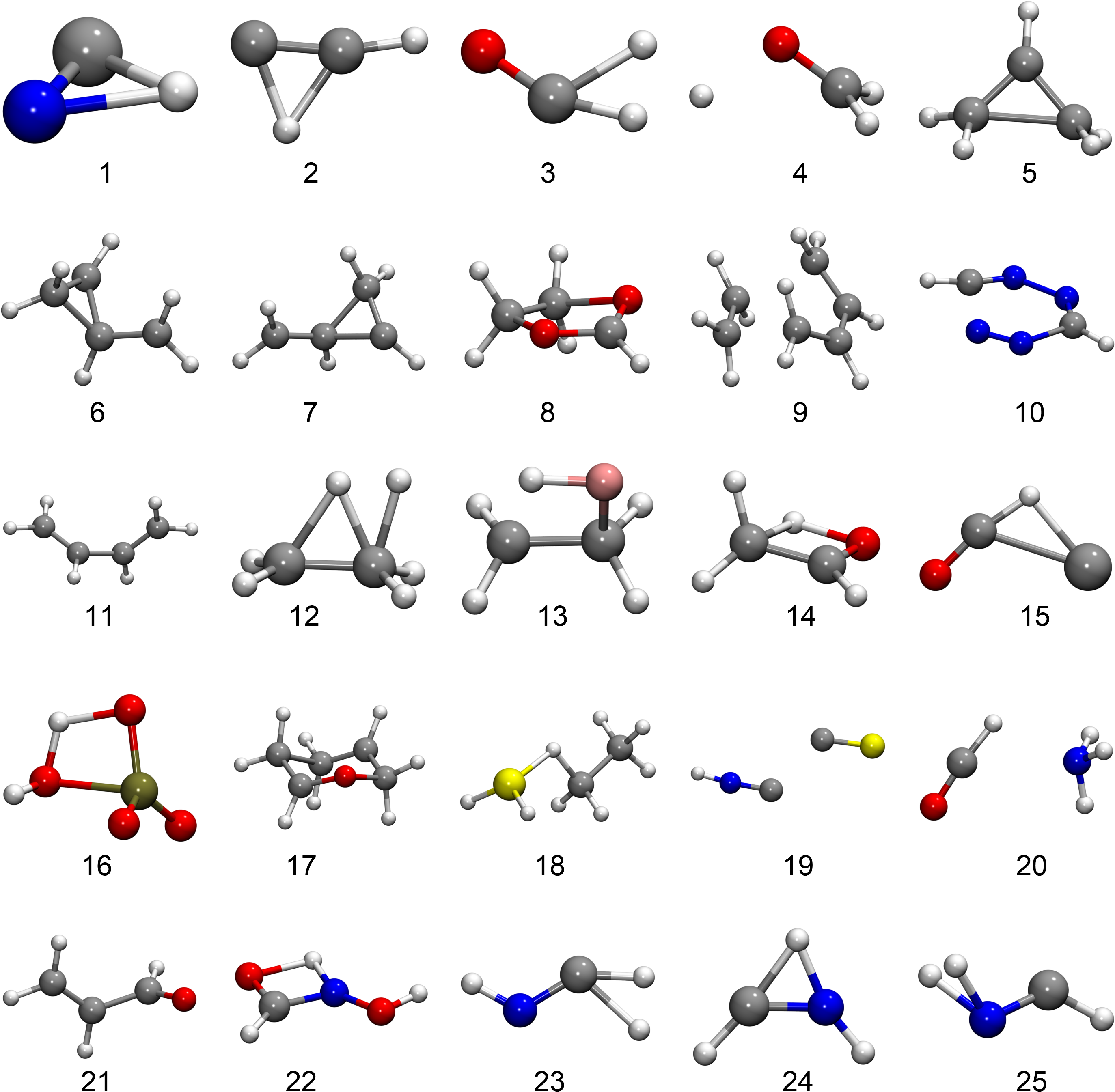} % 8cm
			\caption{All TSs for the Baker test set.
			}
			\label{fig::BakerTS}
		\end{center}
	\end{figure}

%%%%%%%%%%%%%%%%%%%%%%%%%%%%%%%%%%%%%%%%%%%%%%%%%%%%%%%%%%%%%%%%%%%%%%%%
	To benchmark our TS search we chose $27$ test systems:
	first of all, $25$ test systems suggested by Baker.\cite{Baker}
	The starting points of the optimizations 
	were chosen following Ref. \citenum{Baker}
	close to a TS on Hartree--Fock level.
	However, we use the semi-empirical AM1\cite{dewar1985development} 
	method for the electronic structure calculations.	
	We plot all the TSs we found 
	in the Baker test set in \figref{fig::BakerTS}.
	These TSs were found by GPRTS except for the
	TS in system $4$, $10$, and $15$ for which we used the
	dimer method. 
	\ad{These exceptions were made since the 
	GPRTS method finds a different transition than implied by the test set
	for system $4$ and $10$.
	For system $15$ GPRTS does not converge.}
	
\begin{figure}%[h!]
		\begin{center}
			\includegraphics[width=8cm]{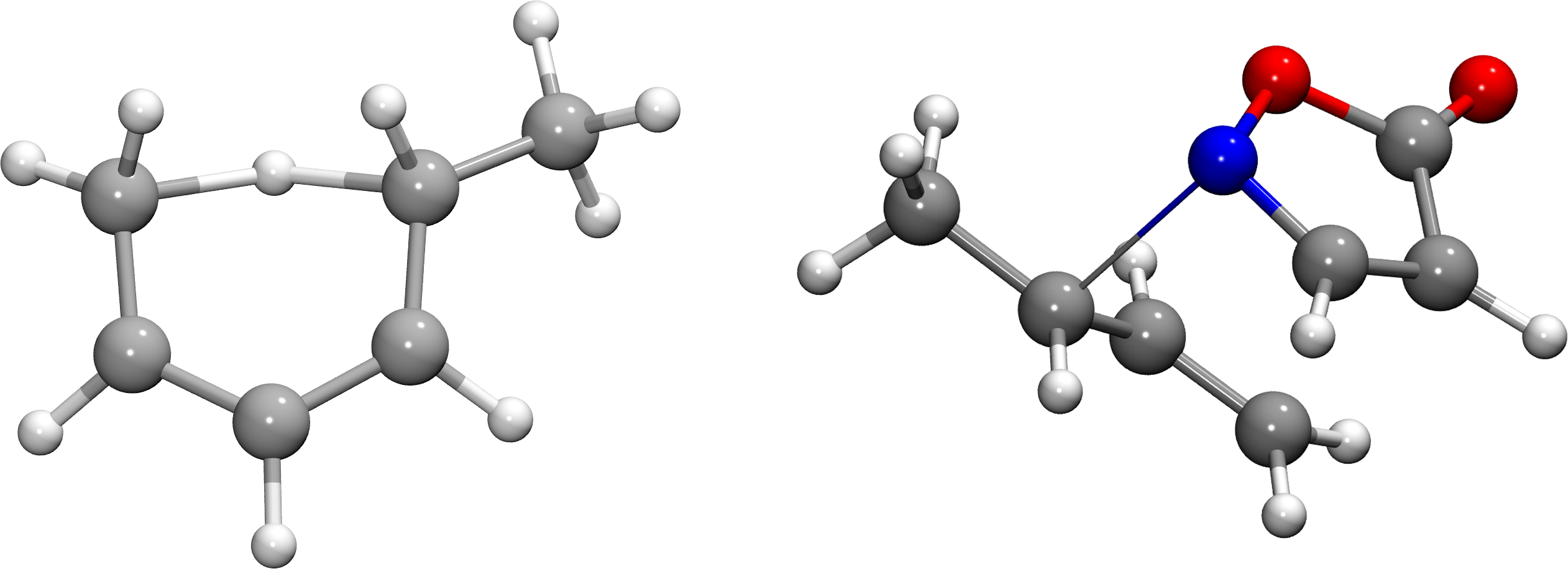} % 8cm
			\caption{TSs found by GPRTS for system 26, the [1,5] H shift of 1,3(Z)-hexadiene
				to 2(E),4(Z)-hexadiene, (left) and system 27, isoxazolinone, (right).}
			\label{fig::hexadiene}
		\end{center}
	\end{figure}

	Furthermore, we chose two test systems on DFT level
	using the BP86 functional\cite{bec88,perdew1986}
	and the 6-31G* basis set.\cite{6-31G_star}
	Firstly, an intramolecular [1,5] H shift of 1,3(Z)-hexadiene
	to 2(E),4(Z)-hexadiene as investigated in Ref. \citenum{dualLevelInst}.
	Secondly, an asymmetric allylation of a simple isoxazolinone
	as investigated in Ref. \citenum{ts_sys_27}, see \figref{fig::hexadiene}.
	The starting points for the TS searches on those two systems
	were chosen by chemical intuition to approximate the real TS.
	
	In the next section %\secref{sec::bench_opt} 
	we compare the new GPR-based
	TS search (GPRTS) against the dimer method
	and the P-RFO method as they are implemented in \dl.
	This is our first benchmark.	
	In the following section %\secref{sec::bench_GPRTSIDPP} 
	we evaluate our GPRPP approach
	as a second benchmark.
	
	Every time the energy is evaluated we also
	evaluate the gradient, a process which will be referred to merely as 
	\emph{energy evaluation} in the following for simplicity.

	\begin{table*}
		\caption{A comparison of GPRTS to dimer and
			P-RFO in our test systems, 
			sorted by the number of dimensions $d$. 
			The number of required energy calculations until convergence is shown,
			as well as the energy difference (Hartree) and 
			RMSD values ({\AA}ng) of the resulting structure
			compared to the structure found by GPRTS.
			Convergence problems of the AM1 method are marked with $\textit{err}$,
			non-convergence after $1000$ energy evaluations is marked
			$\text{nc}$  for \emph{not converged}.
		}	
		{\setlength{\extrarowheight}{2pt}
			\bgroup
			\centering
\begin{tabularx}{\textwidth}{|Y|YYY|S[table-format=+2.2e+2]S[table-format=+2.2e+2]|S[table-format=+1.2e+2]S[table-format=+1.2e+2]|c|}
					\hline
					& \multicolumn{3}{c|}{Energy evaluations}&
					\multicolumn{2}{c|}{Energy differences}&
					\multicolumn{2}{c|}{RMSD values}&\\ \hline
					$d$ & 
					GPRTS & 
					dimer & 
					\mbox{P-RFO} & 					
					\multicolumn{1}{c}{dimer} & 
					\multicolumn{1}{c|}{P-RFO} & 
					\multicolumn{1}{c}{dimer} & 
					\multicolumn{1}{c|}{P-RFO} & 	
					\multicolumn{1}{c|}{ID} \\ 
					\hline	\hline	
$57$ & $87$ & $218$ & $193$ & 1.00e-7 & -2.30e-6 & 1.21e-2 & 5.01e-3 & $27$ \\ \hline
$48$ & $79$ & $155$ & $372$ & -5.00e-7 & -4.11e-2 & 3.54e-3 & 9.84e-1 & $26$ \\ \hline
$48$ & $47$ & $171$ & $\text{nc}$ & 9.97e-7 &   & 2.24e-3 &   & $9$ \\ \hline
$42$ & $34$ & $83$ & $130$ & 1.79e-7 & -1.86e-6 & 3.83e-3 & 3.02e-3 & $17$ \\ \hline
$33$ & $101$ & $288$ & $232$ & 3.96e-8 & 3.33e-5 & 6.30e-3 & 7.12e-2 & $18$ \\ \hline
$30$ & $47$ & $106$ & $175$ & -5.70e-3 & -5.70e-3 & 1.23e-1 & 1.23e-1 & $6$ \\ \hline
$30$ & $41$ & $100$ & $507$ & 1.21e-7 & -3.97e-2 & 1.81e-3 & 5.80e-1 & $7$ \\ \hline
$30$ & $28$ & $63$ & $83$ & 2.49e-8 & -3.21e-8 & 1.69e-3 & 2.87e-4 & $8$ \\ \hline
$30$ & $49$ & $\text{err}$ & $71$ &   &-3.05e-3 &   &7.33e-1 & $11$ \\ \hline
$24$ & $25$ & $59$ & $85$ &-8.90e-9 &-1.60e-7 &2.00e-3 &6.64e-4 & $5$ \\ \hline
$24$ & $25$ & $224$ & $151$ &-1.45e-3 &-1.16e-1 &1.03e-1 &2.29e-1 & $10$ \\ \hline
$24$ & $22$ & $53$ & $68$ &9.42e-7 &1.50e-9 &2.86e-3 &1.81e-4 & $12$ \\ \hline
$24$ & $31$ & $88$ & $73$ &5.80e-7 &-9.05e-9 &2.56e-3 &8.39e-5 & $13$ \\ \hline
$24$ & $100$ & $\text{err}$ & $55$ &   &-2.81e-7 &   &4.40e-3 & $21$ \\ \hline
$21$ & $34$ & $76$ & $80$ &4.18e-8 & 1.07e-8  &4.21e-1 &2.36e-4 & $14$ \\ \hline
$21$ & $31$ & $67$ & $79$ &4.33e-8 &9.00e-10 &2.05e-3 &5.14e-5 & $16$ \\ \hline
$21$ & $24$ & $211$ & $142$ &5.15e-8 &-1.06e-7 &5.24e-3 &3.04e-1 & $20$ \\ \hline
$21$ & $18$ & $44$ & $174$ &7.39e-8 &-1.55e-3 &1.99e-3 &2.71e-1 & $22$ \\ \hline
$15$ & $36$ & $152$ & $66$ &7.97e-2 &1.80e-10 &2.25e-1 &3.18e-5 & $4$ \\ \hline
$15$ & $22$ & $69$ & $44$ &-1.66e-8 &3.35e-6 &2.05e-3 &7.81e-3 & $19$ \\ \hline
$15$ & $16$ & $38$ & $56$ &2.30e-7 &1.06e-8 &2.53e-3 &2.20e-4 & $23$ \\ \hline
$15$ & $19$ & $58$ & $114$ &-5.80e-9 &-1.75e-8 &2.42e-3 &3.92e-4 & $24$ \\ \hline
$15$ & $23$ & $45$ & $59$ &5.00e-8 &1.51e-7 &2.38e-3 &1.11e-3 & $25$ \\ \hline
$12$ & $14$ & $34$ & $35$ &1.00e-9 &6.40e-9 &2.91e-3 &6.77e-5 & $2$ \\ \hline
$12$ & $16$ & $31$ & $127$ &3.41e-9 &-9.43e-2 &2.61e-3 &1.15e0 & $3$ \\ \hline
$12$ & $\text{err}$ & $59$ & $\text{err}$ &   &   &   &   & $15$ \\ \hline
$9$ & $18$ & $39$ & $30$ &1.60e-9 &-1.80e-9 &3.03e-3 &4.03e-5 & $1$ \\ \hline
\end{tabularx}
			\egroup
		}
		\label{tab::nStepsBM1}
	\end{table*}

\paragraph{Benchmarking the optimization.\label{sec::bench_opt}}
	We optimized the TSs of our test set using the new GPRTS
	optimizer and also the established dimer method
	and the P-RFO method in \dl.
	We compare the number of energy evaluations that all three
	methods need until convergence in \tabref{tab::nStepsBM1}.
	Note that analytic Hessians are available for systems $26$ and $27$.
	P-RFO uses $8$ ($4$) analytic Hessians for system $26$ ($27$).
	Those are not counted as energy evaluations.
	In the other systems the Hessians are calculated \ad{using
	central difference approximation via gradients.}
	The required $2d$ gradients are counted as energy
	evaluations in the table.
	\ad{Note that every
	\emph{energy evaluation} in our tables always refers
	to evaluating the energy and the gradient of the PES.}
	We chose a maximum step size of $s_{\text{max}}=\num{0.3}$
	and a tolerance $\delta=\num{3e-4}$ \ad{for all optimizations.}
	
	The P-RFO method calculates the Hessian at the first
	point and then only every $50$ following steps.
	All other Hessians are inferred via 
	the update mechanism by Bofill. \cite{Bofill} 
	
	We see that the GPRTS method generally requires fewer energy evaluations
	than the other methods and has the least convergence problems.
	In \tabref{tab::nStepsBM1} we also show the energy differences
	(the energy from the respective method minus the energy from GPRTS)
	and root mean squared deviations (RMSD)
	of the found TSs, compared to the TSs found by GPRTS.
	Looking at the specific systems with relevant deviations for the found TSs
	we observe the following differences, compare \figref{fig::BakerTS}:
\begin{itemize}
	\item System 3: P-RFO finds a different TS in which $\ce{H2}$ is completely abstracted. 
					Its energy is $9.34\times 10^{-2}$ Hartree lower than that of the TS found by GPRTS.
	\item System 4: GPRTS and P-RFO yield a smaller distance of the $\ce{H}$-Atom to the remaining $\ce{O}$ atom.
					which corresponds to a different transition than intended: a rotation of the $\ce{OH}$ group.
	\item System 6: GPRTS finds a different angle H--C--C at the carbene end.
	\item System 7: P-RFO finds an opening of the ring structure which remains closed in the other cases.
	\item System 10: GPRTS finds a TS in that $\ce{N2}$ is abstracted,
					the dimer method finds the depicted opening of the ring. 
					The opening of the ring is also found via GPRTS
					if one uses a smaller step size.
					P-RFO finds a closed ring structure as the TS.
	\item System 11: P-RFO finds a more planar structure.
	\item System 14: The dimer method finds a mirror image of the TS found by the other methods, thus the high RMSD.
	\item System 20: P-RFO finds a different orientation of $\ce{NH3}$, 
					rotated relative to $\ce{HCO}$, and also with a slightly different distance.
	\item System 22: P-RFO finds a different angle of the attached $\ce{OH}$ group. 
					The resulting molecule is not planar. 
					\ad{The resulting structures of GPRTS and the dimer method are planar.}
	\item System 26: P-RFO does not find the $\ce{H}$-transfer but an opening of the ring. 
\end{itemize}

	\begin{figure}
		\centering
		\begin{center}
			\includegraphics[width=\linewidth]{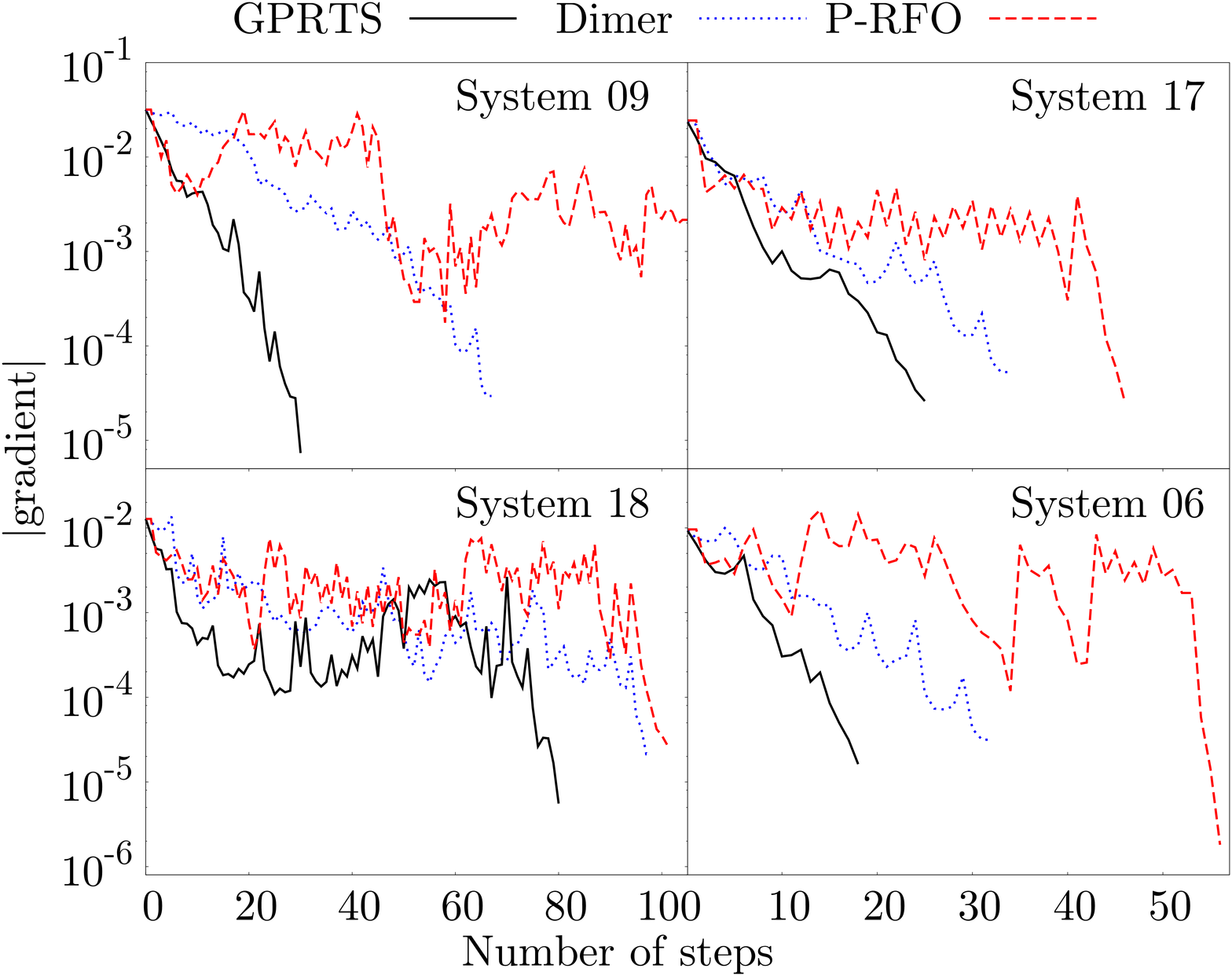}
			\caption{
				The Euclidean norm of the gradient with respect to the
				number of steps on the PES taken by GPRTS, the dimer method,
				and P-RFO in the four biggest systems
				in the Baker test set.
				We explicitly excluded the steps needed to optimize
				the minimum mode/calculate the Hessian.
				We truncated the plot for P-RFO in system $09$ since it
				does not converge.
			}
			\label{fig::convOrder_baker}
		\end{center}
	\end{figure}
	
	\begin{figure}
		\begin{center}
			\includegraphics[width=\linewidth]{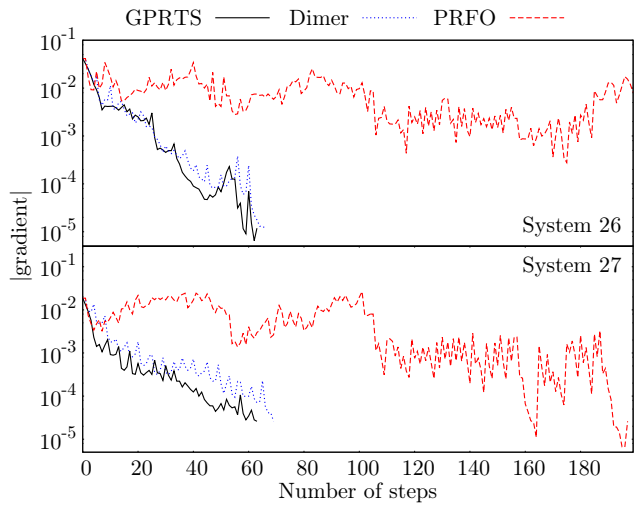}
			\caption{
				The Euclidean norm of the gradient with respect to the
				number of steps on the PES taken by GPRTS, the dimer method,
				and P-RFO in the systems $26$ and $27$.
				We explicitly excluded the steps needed to optimize
				the minimum mode/calculate the Hessian.
				We truncated the plot for P-RFO in system $26$ since
				it does not converge to the same SP as the other methods.
			}
			\label{fig::convOrder_jansystem}
		\end{center}
	\end{figure}
	
	Furthermore, we also compare the evolution of the residual error of %history of the error of 
	GPRTS, P-RFO, and the dimer method during the TS search for the four largest
	test systems in the Baker test set in \figref{fig::convOrder_baker} and
	for systems $26$ and $27$ in \figref{fig::convOrder_jansystem}.
	In both plots we only count the number of steps on the PES
	that are taken along the optimization runs. The energy
	evaluations performed to converge the transition mode, to
    rotate the dimer, and to calculate the Hessian in P-RFO are excluded.
	We see a significant speedup with the new GPRTS optimizer.
	In addition, the new method needs also fewer gradients 
	than the dimer method to approximate
	the Hessian appropriately.
	In our optimizer we often see jumps of 
	the gradient norm at around $50$ steps.
	This is where the multi-level approach creates the first
    separate GP.
	Therefore, we include a possibility to increase 
	the number of training points in the last level $GP_0$ in the code
	to get better performance in high-precision calculations.
	We do not make use of this possibility in all test systems
	presented in the paper.
	
\paragraph{Benchmarking GPRPP.\label{sec::bench_GPRTSIDPP}}
	As described in the previous section about finding a starting point via IDPP %\secref{sec:: GPRPP} 
	we use an approximated NEB path
	between two known minima to find a good starting point for the TS search.
	To prepare the input for our test systems we proceeded as follows: %perform the benchmark in the following way:
\begin{itemize}
	\item We start from the TSs shown in \figref{fig::BakerTS} and \figref{fig::hexadiene}.
	\item We performed IRC path searches to find the two minima connected by the TS. %that the TS is connecting.
	\item The geometries of these minima are used as starting points for GPRPP. %are fed to \dl to optimize
		  %a NEB in the IDPP.
	\item Additionally, GPRTS, dimer and P-RFO searches are started from $\vec{x}^\text{IDPP}_\text{best}$. 
		  In GPRPP, the points on the NEB path are included in the GPR, 
		  while in GPRTS they are excluded. 
\end{itemize}

	In the benchmarks reported in \tabref{tab::nStepsBM2} the additional points to find $\vec{x}^\text{IDPP}_\text{best}$
	are included in the number of steps that the  GPRPP optimizer needs, but
	not in the numbers for the other methods.
	Our approximated NEB path consists of $10$ images for all test cases.
	The number of additional energy evaluations that GPRPP needs to find
	$\vec{x}^\text{IDPP}_\text{best}$ is $2$ or $3$ for all test systems.
	We set the same maximum step size and tolerance value
	as in the previous section. %\secref{sec::bench_opt}.
	We compare the number of energy evaluations all methods
	need until convergence in \tabref{tab::nStepsBM2}.
	We also show the energy differences 
	(the energy from the respective method
	minus the energy from GPRPP) and RMSD values compared to GPRPP.
	Note that analytic Hessians are available for systems $26$ and $27$.
	P-RFO uses \ad{$2$} analytic Hessians for system $26$
	\ad{and $5$ for system $27$.}
	Those are not counted as energy evaluations.
	In the other systems the Hessians are calculated via finite 
	differences and the required $2d$ gradients are counted as energy
	evaluations in the table.
	We observe that GPRPP and GPRTS yield
	consistently good results.
	Comparing GPRPP and GPRTS we see no clear tendency
	whether it is useful to include the points that lead to
	$\vec{v}_{\text{best}}$ in order to improve the speed of the convergence.
	On the other hand, we have one system that only converges when
	these points are included. 
	We conclude that the stability might be improved
	using the GPRPP method with the additional training points
	in comparison to pure GPRTS.
	This will strongly depend on the choice of the starting point for GPRTS.

\begin{sidewaystable*}%[h]
	\centering	
	\caption{
		A Comparison of GPRPP to GPRTS, dimer, and P-RFO in our test systems, 
		sorted by the number of dimensions $d$.
		The number of required energy calculations until convergence is shown,
		as well as the energy difference (Hartree) and 
		RMSD values ({\AA}ng) of the resulting structure
		compared to the structure found by GPRPP.
		Convergence problems of the AM1 method are marked with $\textit{err}$,
		non-convergence after $1000$ energy evaluations is marked
		$\text{nc}$  for \emph{not converged}.
		}		
		\begin{tabularx}{\linewidth}{|c|YYYY|S[table-format=+2.2e+2]S[table-format=+2.2e+2]S[table-format=+2.2e+2]|S[table-format=+1.2e+2]S[table-format=+1.2e+2]S[table-format=+1.2e+2]|c|}
	\hline
	& \multicolumn{4}{c|}{Energy evaluations}&
	\multicolumn{3}{c|}{Energy differences}&
	\multicolumn{3}{c|}{RMSD values}&\multicolumn{1}{c|}{ }\\ \hline%\cline{2-5}\cline{6-8}\cline{9-11}
	$d$ & 
	GPRPP & 
	 GPRTS & 
	 \multicolumn{1}{c}{dimer} & 
	 \multicolumn{1}{c|}{P-RFO} & 
	 \multicolumn{1}{c}{GPRTS} & 
	 \multicolumn{1}{c}{dimer} & 
	 \multicolumn{1}{c|}{P-RFO} & 
	 \multicolumn{1}{c}{GPRTS} & 
	 \multicolumn{1}{c}{dimer} & 
	 \multicolumn{1}{c|}{PRFO} & 
	 ID \\ \hline \hline
$57$ & $87$ & $111$ & $339$ & $210$ & -4.00e-7 & 1.40e-6 & -3.30e-6 & 9.06e-3 & 1.25e-2 & 7.02e-1 & $27$ \\ \hline
$48$ & $54$ & $52$ & $119$ & $59$ & 1.40e-6 & 9.00e-7 & -2.00e-7 & 1.34e-2 & 1.27e-2 & 4.75e-3 & $26$ \\ \hline
$48$ & $40$ & $52$ & $\text{nc}$ & $\text{nc}$ & -1.51e-7 & & & 6.21e-4 & & & $9$ \\ \hline
$42$ & $88$ & $115$ & $\text{err}$ & $225$ & 2.84e-7 & & -4.35e-7 & 6.88e-4 & & 1.17e-3 & $17$ \\ \hline
$33$ & $46$ & $52$ & $151$ & $190$ & -8.13e-5 & -9.10e-5 & -9.10e-5 & 8.70e-2 & 1.28e-1 & 1.28e-1 & $18$ \\ \hline
$30$ & $41$ & $47$ & $121$ & $174$ & -1.21e-6 & -9.49e-7 & -1.23e-6 & 2.54e-3 & 2.30e-3 & 2.55e-3 & $6$ \\ \hline
$30$ & $39$ & $40$ & $114$ & $175$ & -1.82e-7 & -8.00e-10 & -2.41e-7 & 9.85e-4 & 1.07e-3 & 1.04e-3 & $7$ \\ \hline
$30$ & $29$ & $28$ & $74$ & $71$ & -5.38e-8 & 4.50e-8 & -9.23e-8 & 5.52e-4 & 8.28e-4 & 4.73e-4 & $8$ \\ \hline
$30$ & $69$ & $65$ & $\text{nc}$ & $69$ & -1.43e-8 & & -2.92e-3 & 2.03e-4 & & 6.93e-1 & $11$ \\ \hline
$24$ & $25$ & $26$ & $60$ & $75$ & 9.87e-8 & 1.31e-7 & -2.04e-8 & 4.77e-4 & 5.93e-4 & 2.61e-4 & $5$ \\ \hline
$24$ & $34$ & $31$ & $\text{err}$ & $86$ & 1.16e-1 & & 5.91e-7 & 2.29e-1 & & 1.34e-3 & $10$ \\ \hline
$24$ & $93$ & $82$ & $\text{err}$ & $152$ & 6.00e-10 & & 1.00e-10 & 4.64e-5 & & 3.56e-5 & $12$ \\ \hline
$24$ & $40$ & $39$ & $122$ & $92$ & -6.73e-9 & 7.23e-8 & 1.48e-7 & 1.18e-4 & 3.67e-4 & 8.80e-4 & $13$ \\ \hline
$24$ & $69$ & $129$ & $\text{err}$ & $55$ & 4.96e-9 & & -2.77e-7 & 2.60e-4 & & 4.49e-3 & $21$ \\ \hline
$21$ & $31$ & $31$ & $68$ & $73$ & 9.04e-8 & 2.84e-7 & -4.93e-9 & 4.21e-1 & 4.22e-1 & 4.22e-1 & $14$ \\ \hline
$21$ & $30$ & $31$ & $75$ & $68$ & 9.16e-7 & 5.02e-8 & 3.80e-9 & 2.13e-3 & 6.52e-4 & 3.48e-4 & $16$ \\ \hline
$21$ & $51$ & $205$ & $\text{nc}$ & $\text{nc}$ & 2.45e-2 & & & 2.99e0 & & & $20$ \\ \hline
$21$ & $35$ & $32$ & $90$ & $76$ & -1.00e-10 & 3.13e-8 & 1.38e-8 & 1.05e-4 & 2.82e-4 & 1.50e-4 & $22$ \\ \hline
$15$ & $87$ & $55$ & $\text{nc}$ & $142$ & 1.43e-3 & & 1.41e-3 & 5.76e-1 & & 5.29e-1 & $4$ \\ \hline
$15$ & $23$ & $22$ & $68$ & $45$ & 0.00e0 & 3.07e-8 & 4.40e-7 & 1.31e-4 & 2.99e-4 & 3.32e-3 & $19$ \\ \hline
$15$ & $35$ & $95$ & $\text{err}$ & $377$ & -1.01e-1 & & -1.01e-1 & 9.84e-1 & & 7.44e-1 & $23$ \\ \hline
$15$ & $23$ & $25$ & $60$ & $63$ & -8.00e-10 & 5.49e-8 & 7.80e-9 & 3.12e-5 & 5.51e-4 & 1.48e-4 & $24$ \\ \hline
$15$ & $20$ & $21$ & $54$ & $56$ & 3.37e-8 & 1.16e-8 & 4.10e-9 & 3.61e-4 & 1.07e-4 & 2.54e-4 & $25$ \\ \hline
$12$ & $16$ & $17$ & $37$ & $35$ & -1.00e-10 & -1.00e-10 & 5.50e-9 & 3.54e-5 & 4.31e-5 & 8.26e-5 & $2$ \\ \hline
$12$ & $29$ & $17$ & $29$ & $66$ & 9.50e-10 & 2.05e-8 & 3.35e-9 & 9.33e-5 & 1.31e-4 & 1.27e-4 & $3$ \\ \hline
$12$ & $28$ & $\text{err}$ & $\text{err}$ & $\text{err}$ & & & & & & & $15$ \\ \hline
$9$ & $20$ & $19$ & $40$ & $37$ & 8.00e-10 & -1.00e-10 & 0.00e0 & 4.47e-5 & 2.45e-5 & 2.03e-6 & $1$ \\ \hline
\end{tabularx}
	\label{tab::nStepsBM2}
\end{sidewaystable*}
%\end{landscape}

	Looking at the energy differences and RMSD values we see a few
	deviations between the results of the different methods
	that we like to discuss, compare \figref{fig::BakerTS}.
\begin{itemize}
	\item System 4:
		The abstracted $\ce{H}$ atom has a different angle and distance in every optimization.
	\item System 10:
		GPRPP and P-RFO found a closed, symmetric ring structure. 
		GPRTS found a similar ring as in the reference structure depicted in \figref{fig::BakerTS}, only with larger distance of the two nitrogen atoms to the
		carbons.
		The dimer method leads to a problem with the AM1 SCF convergence.
	\item System 11:
		P-RFO finds a more planar structure.
	\item System 12:
		The dimer method does not converge. All other methods yield the same TS,
		in which the two hydrogen atoms that get abstracted are moved more towards one of the carbon atoms.
	\item System 14:
		Dimer, GPRTS and P-RFO all find the mirror image of the TS found by GPRPP. 
	\item System 15:
		GPRPP converges to the correct SP but the suggested starting point
		leads all other methods to a configuration for which the AM1 SCF cycles do not converge anymore.
	\item System 18:
		Dimer and P-RFO	yield slightly different angles between the atoms.
	\item System 20:
		GPRTS yields unrealistic large distances of the two molecules.
		The other two methods do not converge. They start unrealistically increasing 
		the distance of the two molecules as well.
	\item System 23:
		GPRPP finds the same TS as depicted in \figref{fig::BakerTS}, only with a different angle of the hydrogen that is attached to the nitrogen. 
		All other methods find too large, and each a different, distance of the $\ce{H2}$ that is abstracted.
	\ad{\item System 27:
		P-RFO finds a different orientation of the ring structure corresponding to a different TS.}

\end{itemize}
	
	The clear superiority of the new optimizer in the test systems indicates
	that the GPR-based representation of the Hessian is very well suited
	for finding the minimum mode.
	The suggested starting points $\vec{x}^\text{IDPP}_\text{best}$ 
	for the TS searches are plausible in most systems.
	Overall the initial guess for the TS search using the IDPP
	seems to be sufficient for our usage in the GPRPP method
	but is generally not advisable for other TS-search algorithms.
	In some test cases it can only be used if one manually
	corrects for chemically unintuitive TS estimates.

\paragraph{Timing.}
\label{sec::timing}
	Our GPR based optimizer has a larger computational overhead
	than the other optimizers.
	Solving the linear system of \eqref{eq::linSystem} and the
	P-RFO runs on the GPR-PES surface
	can be highly time consuming.
	With the multi-level approach we limit
	the computational effort of solving the linear
	system. 
	Also we parallelized the evaluation of the Hessian matrix,
	needed for P-RFO, with OpenMP.
	We look at the timing in the two DFT-based optimization runs of 
	our first benchmark %\secref{sec::bench_opt}
	in \tabref{tab::timingBM1}.

\begin{table}[h]
	\centering
	\caption{Total time in the GPRTS benchmark.}
	{\setlength{\tabcolsep}{2pt}
		\begin{tabularx}{\linewidth}{cYYY}
		\hline \hline
		\multicolumn{1}{c}{}& \multicolumn{3}{c}{Time [seconds] (\% used by optimizer)} \\ \cline{2-4}
		ID & GPRTS & dimer & P-RFO \\ \hline
		26 & 461 (5\%) & 810 (0\%) & 3665 (0\%) \\		
		27 & 419 (15\%) & 891 (0\%) & 1374 (0\%) \\
		\hline \hline
		\end{tabularx}		
	}
	\label{tab::timingBM1}
\end{table}

The fraction of time spent on the GPRTS optimizer
can vary because the P-RFO optimization
on the GPR-PES converges faster or slower, depending on the system.
The time spent on the runs using the dimer method and P-RFO is almost exclusively 
consumed by the energy evaluations along the runs.

We also have a look 
at the timing of the runs of our second benchmark %\secref{sec::bench_GPRTSIDPP} 
in \tabref{tab::timingBM2}. %\\[2ex]
\ad{Note that P-RFO converges to a different TS in system $27$ and this
	number is not comparable.}

\begin{table}[h]
	\centering
	\caption{Total time in the GPRPP benchmark.}
	{
	\setlength{\tabcolsep}{2pt}
		\begin{tabularx}{\linewidth}{cYYYY}
			\hline \hline
			\multicolumn{1}{c}{}& \multicolumn{4}{c}{Time [seconds] (\% used on optimizer)} \\ \cline{2-5}
			ID & GPRPP & GPRTS & dimer & P-RFO \\ \hline
			26 & 366 (5\%) & 353 (2\%) & 886 (0\%) & 520 (0\%) \\		
			27 & 1041 (5\%) & 1473 (6\%) & 3608 (0\%) & 4129 (0\%)\\
			\hline \hline
		\end{tabularx}			
	}
	\label{tab::timingBM2}
\end{table}

	The presented timings were done on an Intel i5-4570 quad-core CPU.
	The DFT calculations were parallelized on all four cores,
	as well as the evaluation of the Hessian of the GPR-PES.
	Overall, we see that the GPR-based optimizer has an overhead that
	 depends on the system. 
	But the overhead is easily compensated
	by the smaller number of energy calculations that have
	to be performed. The overall time in test systems $26$ and $27$
	is reduced by a factor of $2$ by our new optimizer.
	The improvement will be more significant
	if the optimization is done with a higher level
	of electronic structure calculations.

%%%%%%%%%%%%%%%%%%%%%%%%%%%%%%%%%%%%%%%%%%%%%%%%%%%%%%%%%%%%%%%%%%%%%%%%
\section{Discussion\label{sec::discussion}}
%%%%%%%%%%%%%%%%%%%%%%%%%%%%%%%%%%%%%%%%%%%%%%%%%%%%%%%%%%%%%%%%%%%%%%%%

	For our optimizations we use Cartesian coordinates.
	In many cases it was proven that one can improve the
	interpolation quality
	of machine learning methods by choosing coordinate systems
	that incorporate translational, rotational, 
	\ad{and permutational} invariance.
	\cite{ramakrishnan2017,hansen2015interaction,
	Behler2016,Bartok2013}
	\ad{More modern descriptors like the Coulomb matrix representation
	or the Bag of Bonds approach increase the number of dimensions
	in the system. Considering the scaling of our optimizer
	this might lead to performance problems.
	It was shown that one can use a Z-Matrix representation 
	in the GPR framework for geometry optimization\cite{Christiansen}.
	This introduces new hyperparameters for the length scales.
	Our algorithms are not capable of handling those.}
%	These would usually imply different characteristic length scales
%	in every dimension. Our algorithms are not capable of handling those.
	For traditional TS optimizers the usage
	of internal coordinate systems \ad{like the Z-matrix} 
	is not generally advisable.
	\cite{BakerHehre, Baker}
	Therefore, it is not clear whether the usage of internal 
	coordinates will improve the performance of the optimizer significantly.
	This has to be studied in future work.
	With this in mind it might also be beneficial to
	try a different covariance function/kernel than the one we used.
	By simple cross-validation \ad{(using a
	set of additional
	test points to validate a reasonable choice of the hyperparameters)}
	on several test systems we saw 
	that the squared exponential 
	covariance function yields generally worse results.
	We did not test other covariance functions/kernels.
	We also made no explicit use of the statistical properties
	that GPR yields. One can optimize the hyperparameters
	via the maximum likelihood estimation\cite{rasmussen2006gaussian} and 
	one can make uncertainty predictions of the predicted
	energies via the variance.
	The former was initially used to explore the space of hyperparameters
	$\sigma_\text{e}$, $\sigma_\text{g}$, and $l$.
	But the values that were chosen for this paper were obtained
	via cross-validation on several test systems.
	\ad{We found the optimization of hyperparameters
	based on the maximum likelihood estimation to yield
	no consistent improvements on the overall performance
	of the optimizations.}
	The use of uncertainty predictions, i.e. the variance,
	could be used to find a maximum step size dynamically.
	Nevertheless, we found no advantage of
	a variance-based approach over using a simple distance 
	measure as the maximum step size.
	
	Since the computational effort of our GPR
	approach scales cubically with the number of dimensions,
	we mainly recommend our optimizer for smaller systems.
	In higher dimensional systems one may also have to 
	increase the number of training points in the last level of
	our multi-level	approach which makes the optimizer less efficient.
	We recommend to increase the number of training points for 
	the last level only for smaller systems for which one 
	wants to do high-precision calculations.
	
	Our algorithm apparently is more sensitive 
	to numerical errors than the dimer method:
	choosing different floating point models for the compiler
	can lead to a different performance of the optimizer.
	This effect can also be observed when using P-RFO. 
	The dimer method seems to be almost immune to this effect.
	The solution of the linear system, the evaluations
	of the GPR-PES, the
	diagonalization of the Hessian, and the P-RFO method
	on the GPR-PES all yield slightly different results
	when using different floating-point models.
	These machine-precision errors can lead to
	varying performance of the optimizer.
	In our tests the performance varies only by 
	a few energy evaluations
	(mostly less than $5$, always less than $20$), 
	to the better or worse.
	In some test cases these variations might be higher,
	also using P-RFO.
	For the benchmarks presented in the paper
	we set no explicit compiler flag and use the
	standard setting for floating point operations
	of Intel's fortran compiler, version $16.0.2$.
	That is, using ``more aggressive optimizations
	on floating-point calculations'', as can be read
	in the developer guide for ifort.
	
	The overall result of our benchmark is very promising.
	The big advantage of the new GPRTS optimizer is twofold:
	firstly, one can get a quite precise representation of 
	the second order information with GPR 
	that seems to be superior 
	to traditional Hessian update mechanisms.
	Secondly, our algorithm is able to do very large steps,
	as part of the overshooting procedures.
	Doing large steps and 
	overshooting the estimated solution
	does not hinder the convergence since the optimizer
	can use that information to improve the predicted
	GPR-PES. Therefore, even bad estimates of the
	next point can lead to an improvement
	in the optimization performance.
	
	Our GPRPP method of finding a starting point for the
	TS search seems to work quite well.
	It is generally not advisable to use the estimated
	starting point in other 
	optimization algorithms.
	But it is sufficiently accurate for our GPRPP method.
	Also the additional points of the NEB in the IDPP
	seem to improve the stability of the optimization
	in many cases.
	If chemical intuition or some other method leads to 
	a starting point very close to the real TS, 
	the pure GPRTS method might still be faster.
        
%%%%%%%%%%%%%%%%%%%%%%%%%%%%%%%%%%%%%%%%%%%%%%%%%%%%%%%%%%%%%%%%%%%%%%%%
\section{Conclusions\label{sec::conclusion}}
%%%%%%%%%%%%%%%%%%%%%%%%%%%%%%%%%%%%%%%%%%%%%%%%%%%%%%%%%%%%%%%%%%%%%%%%
	We presented a new black box optimizer 
	to find SPs on energy surfaces based on GPR.
	Only a maximum step size has to be set manually by the user.
	It outperforms both well established methods (dimer and P-RFO).
	The speedup in the presented test systems is significant and
	will be further increased when using higher level theory
	for the electronic structure calculations.
	We also presented an automated way of finding 
	a starting geometry for the
	TS search using the reactant and product geometries.
	We advise to use this approach for systems 
	in which the two minima are known and the 
	estimate of the TS is not straightforward.
	In the presented test systems the method 
	is very stable and fast.
	
%%%%%%%%%%%%%%%%%%%%%%%%%%%%%%%%%%%%%%%%%%%%%%%%%%%%%%%%%%%%%%%%%%%%%%%%	
\section*{Acknowledgments}
	We thank Bernard Haasdonk for stimulating discussions. 
	This work was financially supported by the
    European Union's Horizon 2020 research and innovation programme
    (grant agreement No. 646717, TUNNELCHEM) and the German Research
    Foundation (DFG) through the Cluster of Excellence in Simulation
    Technology (EXC 310/2) at the University of Stuttgart.
%%%%%%%%%%%%%%%%%%%%%%%%%%%%%%%%%%%%%%%%%%%%%%%%%%%%%%%%%%%%%%%%%%%%%%%%	
	
\bibliography{literature}	

\providecommand{\latin}[1]{#1}
\providecommand*\mcitethebibliography{\thebibliography}
\csname @ifundefined\endcsname{endmcitethebibliography}
  {\let\endmcitethebibliography\endthebibliography}{}
\begin{mcitethebibliography}{41}
\providecommand*\natexlab[1]{#1}
\providecommand*\mciteSetBstSublistMode[1]{}
\providecommand*\mciteSetBstMaxWidthForm[2]{}
\providecommand*\mciteBstWouldAddEndPuncttrue
  {\def\EndOfBibitem{\unskip.}}
\providecommand*\mciteBstWouldAddEndPunctfalse
  {\let\EndOfBibitem\relax}
\providecommand*\mciteSetBstMidEndSepPunct[3]{}
\providecommand*\mciteSetBstSublistLabelBeginEnd[3]{}
\providecommand*\EndOfBibitem{}
\mciteSetBstSublistMode{f}
\mciteSetBstMaxWidthForm{subitem}{(\alph{mcitesubitemcount})}
\mciteSetBstSublistLabelBeginEnd
  {\mcitemaxwidthsubitemform\space}
  {\relax}
  {\relax}

\bibitem[Baker(1986)]{Baker_P_RFO}
Baker,~J. An algorithm for the location of transition states. \emph{J. Comput.
  Chem.} \textbf{1986}, \emph{7}, 385--395\relax
\mciteBstWouldAddEndPuncttrue
\mciteSetBstMidEndSepPunct{\mcitedefaultmidpunct}
{\mcitedefaultendpunct}{\mcitedefaultseppunct}\relax
\EndOfBibitem
\bibitem[Banerjee \latin{et~al.}(1985)Banerjee, Adams, Simons, and
  Shepard]{RFO}
Banerjee,~A.; Adams,~N.; Simons,~J.; Shepard,~R. Search for stationary points
  on surfaces. \emph{J. Phys. Chem.} \textbf{1985}, \emph{89}, 52--57\relax
\mciteBstWouldAddEndPuncttrue
\mciteSetBstMidEndSepPunct{\mcitedefaultmidpunct}
{\mcitedefaultendpunct}{\mcitedefaultseppunct}\relax
\EndOfBibitem
\bibitem[Plasencia~Guti\'{e}rrez \latin{et~al.}(2017)Plasencia~Guti\'{e}rrez,
  Arg\'{a}ez, and J\'{o}nsson]{MinModeFollowingJonsson1}
Plasencia~Guti\'{e}rrez,~M.; Arg\'{a}ez,~C.; J\'{o}nsson,~H. Improved Minimum
  Mode Following Method for Finding First Order Saddle Points. \emph{J. Chem.
  Theory Comput.} \textbf{2017}, \emph{13}, 125--134\relax
\mciteBstWouldAddEndPuncttrue
\mciteSetBstMidEndSepPunct{\mcitedefaultmidpunct}
{\mcitedefaultendpunct}{\mcitedefaultseppunct}\relax
\EndOfBibitem
\bibitem[Henkelman and J\'{o}nsson(1999)Henkelman, and J\'{o}nsson]{dimer}
Henkelman,~G.; J\'{o}nsson,~H. A dimer method for finding saddle points on high
  dimensional potential surfaces using only first derivatives. \emph{J. Chem.
  Phys.} \textbf{1999}, \emph{111}, 7010--7022\relax
\mciteBstWouldAddEndPuncttrue
\mciteSetBstMidEndSepPunct{\mcitedefaultmidpunct}
{\mcitedefaultendpunct}{\mcitedefaultseppunct}\relax
\EndOfBibitem
\bibitem[Lanczos(1950)]{Lanczos1950}
Lanczos,~C. {An iteration method for the solution of the eigenvalue problem of
  linear differential and integral operators}. \emph{J. Res. Natl. Bur. Stand.
  B} \textbf{1950}, \emph{45}, 255--282\relax
\mciteBstWouldAddEndPuncttrue
\mciteSetBstMidEndSepPunct{\mcitedefaultmidpunct}
{\mcitedefaultendpunct}{\mcitedefaultseppunct}\relax
\EndOfBibitem
\bibitem[Zeng \latin{et~al.}(2014)Zeng, Xiao, and
  Henkelman]{compareDimerLanczos}
Zeng,~Y.; Xiao,~P.; Henkelman,~G. Unification of algorithms for minimum mode
  optimization. \emph{J. Chem. Phys.} \textbf{2014}, \emph{140}, 044115\relax
\mciteBstWouldAddEndPuncttrue
\mciteSetBstMidEndSepPunct{\mcitedefaultmidpunct}
{\mcitedefaultendpunct}{\mcitedefaultseppunct}\relax
\EndOfBibitem
\bibitem[Heyden \latin{et~al.}(2005)Heyden, Bell, and Keil]{compareDimerP_RFO}
Heyden,~A.; Bell,~A.~T.; Keil,~F.~J. Efficient methods for finding transition
  states in chemical reactions: Comparison of improved dimer method and
  partitioned rational function optimization method. \emph{J. Chem. Phys.}
  \textbf{2005}, \emph{123}, 224101\relax
\mciteBstWouldAddEndPuncttrue
\mciteSetBstMidEndSepPunct{\mcitedefaultmidpunct}
{\mcitedefaultendpunct}{\mcitedefaultseppunct}\relax
\EndOfBibitem
\bibitem[Rasmussen and Williams(2006)Rasmussen, and
  Williams]{rasmussen2006gaussian}
Rasmussen,~C.~E.; Williams,~C.~K. \emph{Gaussian processes for machine
  learning}; MIT press Cambridge, 2006; Vol.~1\relax
\mciteBstWouldAddEndPuncttrue
\mciteSetBstMidEndSepPunct{\mcitedefaultmidpunct}
{\mcitedefaultendpunct}{\mcitedefaultseppunct}\relax
\EndOfBibitem
\bibitem[Alborzpour \latin{et~al.}(2016)Alborzpour, Tew, and
  Habershon]{Habershon_Tew}
Alborzpour,~J.~P.; Tew,~D.~P.; Habershon,~S. Efficient and accurate evaluation
  of potential energy matrix elements for quantum dynamics using Gaussian
  process regression. \emph{J. Chem. Phys.} \textbf{2016}, \emph{145},
  174112\relax
\mciteBstWouldAddEndPuncttrue
\mciteSetBstMidEndSepPunct{\mcitedefaultmidpunct}
{\mcitedefaultendpunct}{\mcitedefaultseppunct}\relax
\EndOfBibitem
\bibitem[Bart\'ok \latin{et~al.}(2013)Bart\'ok, Kondor, and
  Cs\'anyi]{Bartok2013}
Bart\'ok,~A.~P.; Kondor,~R.; Cs\'anyi,~G. On representing chemical
  environments. \emph{Phys. Rev. B} \textbf{2013}, \emph{87}, 184115\relax
\mciteBstWouldAddEndPuncttrue
\mciteSetBstMidEndSepPunct{\mcitedefaultmidpunct}
{\mcitedefaultendpunct}{\mcitedefaultseppunct}\relax
\EndOfBibitem
\bibitem[Ramakrishnan and von Lilienfeld(2017)Ramakrishnan, and von
  Lilienfeld]{ramakrishnan2017}
Ramakrishnan,~R.; von Lilienfeld,~O.~A. \emph{Reviews in Computational
  Chemistry}; JWS, 2017; pp 225--256\relax
\mciteBstWouldAddEndPuncttrue
\mciteSetBstMidEndSepPunct{\mcitedefaultmidpunct}
{\mcitedefaultendpunct}{\mcitedefaultseppunct}\relax
\EndOfBibitem
\bibitem[Mills and Popelier(2011)Mills, and Popelier]{IntramolMultipolKriging}
Mills,~M.~J.; Popelier,~P.~L. Intramolecular polarisable multipolar
  electrostatics from the machine learning method Kriging. \emph{Comput. Theor.
  Chem.} \textbf{2011}, \emph{975}, 42 -- 51\relax
\mciteBstWouldAddEndPuncttrue
\mciteSetBstMidEndSepPunct{\mcitedefaultmidpunct}
{\mcitedefaultendpunct}{\mcitedefaultseppunct}\relax
\EndOfBibitem
\bibitem[Handley \latin{et~al.}(2009)Handley, Hawe, Kell, and
  Popelier]{polWaterKriging}
Handley,~C.~M.; Hawe,~G.~I.; Kell,~D.~B.; Popelier,~P. L.~A. Optimal
  construction of a fast and accurate polarisable water potential based on
  multipole moments trained by machine learning. \emph{Phys. Chem. Chem. Phys.}
  \textbf{2009}, \emph{11}, 6365--6376\relax
\mciteBstWouldAddEndPuncttrue
\mciteSetBstMidEndSepPunct{\mcitedefaultmidpunct}
{\mcitedefaultendpunct}{\mcitedefaultseppunct}\relax
\EndOfBibitem
\bibitem[Fletcher \latin{et~al.}(2014)Fletcher, Kandathil, and
  Popelier]{PredKinEnergyOfCoordsKriging}
Fletcher,~T.~L.; Kandathil,~S.~M.; Popelier,~P. L.~A. The prediction of atomic
  kinetic energies from coordinates of surrounding atoms using kriging machine
  learning. \emph{Theor. Chem. Acc.} \textbf{2014}, \emph{133}, 1499\relax
\mciteBstWouldAddEndPuncttrue
\mciteSetBstMidEndSepPunct{\mcitedefaultmidpunct}
{\mcitedefaultendpunct}{\mcitedefaultseppunct}\relax
\EndOfBibitem
\bibitem[Ramakrishnan and von Lilienfeld(2015)Ramakrishnan, and von
  Lilienfeld]{Ramakrishnan2015}
Ramakrishnan,~R.; von Lilienfeld,~O.~A. Many molecular properties from one
  kernel in chemical space. \emph{CHIMIA} \textbf{2015}, \emph{69},
  182--186\relax
\mciteBstWouldAddEndPuncttrue
\mciteSetBstMidEndSepPunct{\mcitedefaultmidpunct}
{\mcitedefaultendpunct}{\mcitedefaultseppunct}\relax
\EndOfBibitem
\bibitem[Hansen \latin{et~al.}(2015)Hansen, Biegler, Ramakrishnan, Pronobis,
  von Lilienfeld, M{\"u}ller, and Tkatchenko]{hansen2015interaction}
Hansen,~K.; Biegler,~F.; Ramakrishnan,~R.; Pronobis,~W.; von Lilienfeld,~O.~A.;
  M{\"u}ller,~K.-R.; Tkatchenko,~A. Machine Learning Predictions of Molecular
  Properties: Accurate Many-Body Potentials and Nonlocality in Chemical Space.
  \emph{J. Phys. Chem. Lett.} \textbf{2015}, \emph{6}, 2326--2331\relax
\mciteBstWouldAddEndPuncttrue
\mciteSetBstMidEndSepPunct{\mcitedefaultmidpunct}
{\mcitedefaultendpunct}{\mcitedefaultseppunct}\relax
\EndOfBibitem
\bibitem[Dral \latin{et~al.}(2017)Dral, Owens, Yurchenko, and
  Thiel]{PESandVibLevels}
Dral,~P.; Owens,~A.; Yurchenko,~S.; Thiel,~W. Structure-based sampling and
  self-correcting machine learning for accurate calculations of potential
  energy surfaces and vibrational levels. \emph{J. Chem. Phys.} \textbf{2017},
  \emph{146}, 244108\relax
\mciteBstWouldAddEndPuncttrue
\mciteSetBstMidEndSepPunct{\mcitedefaultmidpunct}
{\mcitedefaultendpunct}{\mcitedefaultseppunct}\relax
\EndOfBibitem
\bibitem[Deringer \latin{et~al.}(2018)Deringer, Bernstein, Bart\'ok, Cliffe,
  Kerber, Marbella, Grey, Elliott, and Cs\'anyi]{AtomStrucAmorphousSilicon}
Deringer,~V.~L.; Bernstein,~N.; Bart\'ok,~A.~P.; Cliffe,~M.~J.; Kerber,~R.~N.;
  Marbella,~L.~E.; Grey,~C.~P.; Elliott,~S.~R.; Cs\'anyi,~G. Realistic
  Atomistic Structure of Amorphous Silicon from Machine-Learning-Driven
  Molecular Dynamics. \emph{J. Phys. Chem. Lett.} \textbf{2018}, \emph{9},
  2879--2885\relax
\mciteBstWouldAddEndPuncttrue
\mciteSetBstMidEndSepPunct{\mcitedefaultmidpunct}
{\mcitedefaultendpunct}{\mcitedefaultseppunct}\relax
\EndOfBibitem
\bibitem[Schmitz and Christiansen(2018)Schmitz, and Christiansen]{Christiansen}
Schmitz,~G.; Christiansen,~O. Gaussian process regression to accelerate
  geometry optimizations relying on numerical differentiation. \emph{J. Chem.
  Phys.} \textbf{2018}, \emph{148}, 241704\relax
\mciteBstWouldAddEndPuncttrue
\mciteSetBstMidEndSepPunct{\mcitedefaultmidpunct}
{\mcitedefaultendpunct}{\mcitedefaultseppunct}\relax
\EndOfBibitem
\bibitem[Denzel and K{\"a}stner(2018)Denzel, and K{\"a}stner]{GPRopt}
Denzel,~A.; K{\"a}stner,~J. Gaussian process regression for geometry
  optimization. \emph{J. Chem. Phys.} \textbf{2018}, \emph{148}, 094114\relax
\mciteBstWouldAddEndPuncttrue
\mciteSetBstMidEndSepPunct{\mcitedefaultmidpunct}
{\mcitedefaultendpunct}{\mcitedefaultseppunct}\relax
\EndOfBibitem
\bibitem[Mills and J\'{o}nsson(1994)Mills, and J\'{o}nsson]{MillsNEB_brief}
Mills,~G.; J\'{o}nsson,~H. Quantum and thermal effects in ${\mathrm{H}}_{2}$
  dissociative adsorption: Evaluation of free energy barriers in
  multidimensional quantum systems. \emph{Phys. Rev. Lett.} \textbf{1994},
  \emph{72}, 1124--1127\relax
\mciteBstWouldAddEndPuncttrue
\mciteSetBstMidEndSepPunct{\mcitedefaultmidpunct}
{\mcitedefaultendpunct}{\mcitedefaultseppunct}\relax
\EndOfBibitem
\bibitem[Henkelman \latin{et~al.}(2000)Henkelman, Uberuaga, and
  J\'{o}nsson]{HenkelmanNEB}
Henkelman,~G.; Uberuaga,~B.~P.; J\'{o}nsson,~H. A climbing image nudged elastic
  band method for finding saddle points and minimum energy paths. \emph{J.
  Chem. Phys.} \textbf{2000}, \emph{113}, 9901--9904\relax
\mciteBstWouldAddEndPuncttrue
\mciteSetBstMidEndSepPunct{\mcitedefaultmidpunct}
{\mcitedefaultendpunct}{\mcitedefaultseppunct}\relax
\EndOfBibitem
\bibitem[Koistinen \latin{et~al.}(2016)Koistinen, Maras, Vehtari, and
  J\'{o}nsson]{GPR_MEP}
Koistinen,~O.-P.; Maras,~E.; Vehtari,~A.; J\'{o}nsson,~H. Minimum energy path
  calculations with Gaussian process regression. \emph{Nanosystems: Phys. Chem.
  Math.} \textbf{2016}, \emph{7}, 925--935\relax
\mciteBstWouldAddEndPuncttrue
\mciteSetBstMidEndSepPunct{\mcitedefaultmidpunct}
{\mcitedefaultendpunct}{\mcitedefaultseppunct}\relax
\EndOfBibitem
\bibitem[Koistinen \latin{et~al.}(2017)Koistinen, Dagbjartsd\'{o}ttir,
  \'{A}sgeirsson, Vehtari, and J\'{o}nsson]{GPRNEB_Jonsson}
Koistinen,~O.-P.; Dagbjartsd\'{o}ttir,~F.~B.; \'{A}sgeirsson,~V.; Vehtari,~A.;
  J\'{o}nsson,~H. Nudged elastic band calculations accelerated with Gaussian
  process regression. \emph{J. Chem. Phys.} \textbf{2017}, \emph{147},
  152720\relax
\mciteBstWouldAddEndPuncttrue
\mciteSetBstMidEndSepPunct{\mcitedefaultmidpunct}
{\mcitedefaultendpunct}{\mcitedefaultseppunct}\relax
\EndOfBibitem
\bibitem[Smidstrup \latin{et~al.}(2014)Smidstrup, Pedersen, Stokbro, and
  J\'{o}nsson]{IDPP}
Smidstrup,~S.; Pedersen,~A.; Stokbro,~K.; J\'{o}nsson,~H. Improved initial
  guess for minimum energy path calculations. \emph{J. Chem. Phys.}
  \textbf{2014}, \emph{140}, 214106\relax
\mciteBstWouldAddEndPuncttrue
\mciteSetBstMidEndSepPunct{\mcitedefaultmidpunct}
{\mcitedefaultendpunct}{\mcitedefaultseppunct}\relax
\EndOfBibitem
\bibitem[K{\"a}stner \latin{et~al.}(2009)K{\"a}stner, Carr, Keal, Thiel,
  Wander, and Sherwood]{dlfind}
K{\"a}stner,~J.; Carr,~J.~M.; Keal,~T.~W.; Thiel,~W.; Wander,~A.; Sherwood,~P.
  {DL-FIND: An Open-Source Geometry Optimizer for Atomistic Simulations}.
  \emph{J. Phys. Chem. A} \textbf{2009}, \emph{113}, 11856--11865\relax
\mciteBstWouldAddEndPuncttrue
\mciteSetBstMidEndSepPunct{\mcitedefaultmidpunct}
{\mcitedefaultendpunct}{\mcitedefaultseppunct}\relax
\EndOfBibitem
\bibitem[K{\"a}stner and Sherwood(2008)K{\"a}stner, and Sherwood]{dimer_dlfind}
K{\"a}stner,~J.; Sherwood,~P. Superlinearly converging dimer method for
  transition state search. \emph{J. Chem. Phys.} \textbf{2008}, \emph{128},
  014106\relax
\mciteBstWouldAddEndPuncttrue
\mciteSetBstMidEndSepPunct{\mcitedefaultmidpunct}
{\mcitedefaultendpunct}{\mcitedefaultseppunct}\relax
\EndOfBibitem
\bibitem[Sherwood \latin{et~al.}(2003)Sherwood, de~Vries, Guest, Schreckenbach,
  Catlow, French, Sokol, Bromley, Thiel, Turner, Billeter, Terstegen, Thiel,
  Kendrick, Rogers, Casci, Watson, King, Karlsen, Sj{\o}voll, Fahmi,
  Sch{\"a}fer, and Lennartz]{SHERWOOD20031}
Sherwood,~P.; de~Vries,~A.~H.; Guest,~M.~F.; Schreckenbach,~G.; Catlow,~C.~A.;
  French,~S.~A.; Sokol,~A.~A.; Bromley,~S.~T.; Thiel,~W.; Turner,~A.~J.
  \latin{et~al.}  QUASI: A general purpose implementation of the QM/MM approach
  and its application to problems in catalysis. \emph{J. Mol. Struct.
  Theochem.} \textbf{2003}, \emph{632}, 1 -- 28\relax
\mciteBstWouldAddEndPuncttrue
\mciteSetBstMidEndSepPunct{\mcitedefaultmidpunct}
{\mcitedefaultendpunct}{\mcitedefaultseppunct}\relax
\EndOfBibitem
\bibitem[Metz \latin{et~al.}(2014)Metz, K{\"a}stner, Sokol, Keal, and
  Sherwood]{ChemshellReview}
Metz,~S.; K{\"a}stner,~J.; Sokol,~A.~A.; Keal,~T.~W.; Sherwood,~P. ChemShell-a
  modular software package for QM/MM simulations. \emph{Wiley Interdiscip. Rev.
  Comput. Mol. Sci.} \textbf{2014}, \emph{4}, 101--110\relax
\mciteBstWouldAddEndPuncttrue
\mciteSetBstMidEndSepPunct{\mcitedefaultmidpunct}
{\mcitedefaultendpunct}{\mcitedefaultseppunct}\relax
\EndOfBibitem
\bibitem[Mat{\'e}rn(2013)]{matern2013spatial}
Mat{\'e}rn,~B. \emph{Spatial variation}; SSBM, 2013; Vol.~36\relax
\mciteBstWouldAddEndPuncttrue
\mciteSetBstMidEndSepPunct{\mcitedefaultmidpunct}
{\mcitedefaultendpunct}{\mcitedefaultseppunct}\relax
\EndOfBibitem
\bibitem[Baker and Chan(1996)Baker, and Chan]{Baker}
Baker,~J.; Chan,~F. The location of transition states: A comparison of
  Cartesian, Z-matrix, and natural internal coordinates. \emph{J. Comput.
  Chem.} \textbf{1996}, \emph{17}, 888--904\relax
\mciteBstWouldAddEndPuncttrue
\mciteSetBstMidEndSepPunct{\mcitedefaultmidpunct}
{\mcitedefaultendpunct}{\mcitedefaultseppunct}\relax
\EndOfBibitem
\bibitem[Dewar \latin{et~al.}(1985)Dewar, Zoebisch, Healy, and
  Stewart]{dewar1985development}
Dewar,~M.~J.; Zoebisch,~E.~G.; Healy,~E.~F.; Stewart,~J.~J. Development and use
  of quantum mechanical molecular models. 76. AM1: a new general purpose
  quantum mechanical molecular model. \emph{J. Am. Chem. Soc.} \textbf{1985},
  \emph{107}, 3902--3909\relax
\mciteBstWouldAddEndPuncttrue
\mciteSetBstMidEndSepPunct{\mcitedefaultmidpunct}
{\mcitedefaultendpunct}{\mcitedefaultseppunct}\relax
\EndOfBibitem
\bibitem[Becke(1988)]{bec88}
Becke,~A. Density-functional exchange-energy approximation with correct
  asymptotic behavior. \emph{Phys. Rev. A} \textbf{1988}, \emph{38},
  3098--3100\relax
\mciteBstWouldAddEndPuncttrue
\mciteSetBstMidEndSepPunct{\mcitedefaultmidpunct}
{\mcitedefaultendpunct}{\mcitedefaultseppunct}\relax
\EndOfBibitem
\bibitem[Perdew(1986)]{perdew1986}
Perdew,~J.~P. Density-functional approximation for the correlation energy of
  the inhomogeneous electron gas. \emph{Phys. Rev. B} \textbf{1986}, \emph{33},
  8822--8824\relax
\mciteBstWouldAddEndPuncttrue
\mciteSetBstMidEndSepPunct{\mcitedefaultmidpunct}
{\mcitedefaultendpunct}{\mcitedefaultseppunct}\relax
\EndOfBibitem
\bibitem[Hariharan and Pople(1973)Hariharan, and Pople]{6-31G_star}
Hariharan,~P.~C.; Pople,~J.~A. The influence of polarization functions on
  molecular orbital hydrogenation energies. \emph{Theor. Chem. Acc.}
  \textbf{1973}, \emph{28}, 213--222\relax
\mciteBstWouldAddEndPuncttrue
\mciteSetBstMidEndSepPunct{\mcitedefaultmidpunct}
{\mcitedefaultendpunct}{\mcitedefaultseppunct}\relax
\EndOfBibitem
\bibitem[Meisner and K{\"a}stner(2018)Meisner, and K{\"a}stner]{dualLevelInst}
Meisner,~J.; K{\"a}stner,~J. Dual-Level Approach to Instanton Theory. \emph{J.
  Chem. Theory Comput.} \textbf{2018}, \emph{14}, 1865--1872\relax
\mciteBstWouldAddEndPuncttrue
\mciteSetBstMidEndSepPunct{\mcitedefaultmidpunct}
{\mcitedefaultendpunct}{\mcitedefaultseppunct}\relax
\EndOfBibitem
\bibitem[Rieckhoff \latin{et~al.}(2017)Rieckhoff, Meisner, K{\"a}stner, Frey,
  and Peters]{ts_sys_27}
Rieckhoff,~S.; Meisner,~J.; K{\"a}stner,~J.; Frey,~W.; Peters,~R. Double
  Regioselective Asymmetric C-Allylation of Isoxazolinones: Iridium-Catalyzed
  N-Allylation Followed by an Aza-Cope Rearrangement. \emph{Angew. Chem. Int.
  Ed.} \textbf{2017}, \emph{57}, 1404--1408\relax
\mciteBstWouldAddEndPuncttrue
\mciteSetBstMidEndSepPunct{\mcitedefaultmidpunct}
{\mcitedefaultendpunct}{\mcitedefaultseppunct}\relax
\EndOfBibitem
\bibitem[Bofill(1994)]{Bofill}
Bofill,~J.~M. Updated Hessian matrix and the restricted step method for
  locating transition structures. \emph{J. Comput. Chem.} \textbf{1994},
  \emph{15}, 1--11\relax
\mciteBstWouldAddEndPuncttrue
\mciteSetBstMidEndSepPunct{\mcitedefaultmidpunct}
{\mcitedefaultendpunct}{\mcitedefaultseppunct}\relax
\EndOfBibitem
\bibitem[Behler(2016)]{Behler2016}
Behler,~J. \emph{J. Chem. Phys.} \textbf{2016}, \emph{145}, 170901\relax
\mciteBstWouldAddEndPuncttrue
\mciteSetBstMidEndSepPunct{\mcitedefaultmidpunct}
{\mcitedefaultendpunct}{\mcitedefaultseppunct}\relax
\EndOfBibitem
\bibitem[Baker and Hehre(1990)Baker, and Hehre]{BakerHehre}
Baker,~J.; Hehre,~W.~J. Geometry optimization in cartesian coordinates: The end
  of the Z-matrix? \emph{J. Comput. Chem.} \textbf{1990}, \emph{12},
  606--610\relax
\mciteBstWouldAddEndPuncttrue
\mciteSetBstMidEndSepPunct{\mcitedefaultmidpunct}
{\mcitedefaultendpunct}{\mcitedefaultseppunct}\relax
\EndOfBibitem
\end{mcitethebibliography}

\newpage

\end{document}